\documentclass[aps,prd,twocolumn,superscriptaddress,showpacs,nofootinbib]{revtex4}

\usepackage[dvips]{graphicx}
\usepackage{ulem}
\usepackage{amsmath}
\usepackage{float}
\def\slash#1{\not\!#1}

\usepackage{color}
\begin{document}


\title{Charmed Baryon $\Lambda_c$ in Nuclear Matter}


\author{Keisuke Ohtani}
\email{ohtani.k@th.phys.titech.ac.jp}
\affiliation{Department of Physics, H-27, Tokyo Institute of Technology, Meguro, Tokyo 152-8551, Japan}

\author{Ken-ji Araki}
\affiliation{Department of Physics, H-27, Tokyo Institute of Technology, Meguro, Tokyo 152-8551, Japan}

\author{Makoto Oka}
\affiliation{Department of Physics, H-27, Tokyo Institute of Technology, Meguro, Tokyo 152-8551, Japan}
\affiliation{Advanced Science Research Center, Japan Atomic Energy Agency, Tokai, Ibaraki 319-1195, Japan}

\date{\today}
\begin{abstract}
Density dependences of the mass and self-energies of $\Lambda_c$ in nuclear matter 
are studied in the parity projected QCD sum rule. 
Effects of nuclear matter are taken into account through the quark and gluon condensates.
It is found that the four-quark condensates give dominant contributions.
As the density dependences of the four-quark condensates are not known well, 
we examine two hypotheses. 
One is based on the factorization hypothesis (F-type) and 
the other is derived from the perturbative chiral quark model (QM-type).
The F-type strongly depends on density, while the QM-type gives a weaker dependence.
It is found that, for the F-type dependence, the energy of $\Lambda_c$ increases 
as the density of nuclear matter grows, that is, $\Lambda_c$ feels repulsion.
On the other hand, the QM-type predicts a weak attraction 
($\sim 20$ MeV at the normal nuclear density)
for $\Lambda_c$ in nuclear matter.
We carry out a similar analysis of the $\Lambda$ hyperon and find that the F-type 
density dependence is too strong to explain the observed binding energy of $\Lambda$ in nuclei.
Thus we conclude that the weak density dependence of the four-quark condensate is more realistic.
The scalar and vector self-energies of $\Lambda_c$ for the QM-type dependence are found to be much smaller 
than those of the light baryons.
\end{abstract}


\maketitle







\section{Introduction}
Heavy hadrons attract much interest because they 
have new properties that are not seen in light hadrons. 
One of the interesting features of the heavy hadron physics is that 
the dynamics of light-quark part can be separated from the 
heavy quark due to the heavy quark spin symmetry \cite{Neubert:1993mb}. 
The spin of heavy quark and that of the rest of the hadron become independent in the 
heavy quark mass limit. 
Therefore, one naively expects that the dynamics of light quarks in a heavy hadron can be 
analyzed more effectively. 
For instance, the heavy baryon consisting of a heavy quark and two light quarks can 
be used in a study of light di-quarks. 
Investigation of the heavy baryon in nuclear matter may help us to 
understand the relation between the partial restoration of chiral symmetry and the 
light di-quark. 

The in-medium modification of the heavy baryon in nuclear matter is also interesting from the 
viewpoint of the heavy baryon-nucleon interaction. 
The N-N and Y-N interactions have been studied both theoretically and experimentally and understood fairly well. 
On the other hand, the properties of the interaction between the charmed (or bottomed) baryon and the nucleon 
are still open issue. 
The existence of $\Lambda_c$ nuclei was discussed about 40 years ago \cite{Dover:1977jw} and subsequently studied 
in Refs.\,\cite{Bhamathi:1981yu,Bando:1981ti,Bando:1983yt,Bando:1985up,Gibson:1983zw,Tsushima:2002cc,Tsushima:2002ua,Tsushima:2002sm}. 
Recently, the $\Lambda_c$-N interaction were reinvestigated by using elaborated models \cite{Liu:2011xc,Maeda:2015hxa} 
and by lattice QCD simulation \cite{Miyamoto:2016hqo}. 

In this study, we investigate the in-medium properties of the $\Lambda_c$ baryon from QCD sum rule. 
We additionally study the properties of the $\Lambda$ hyperon and $\Lambda_b$ and compare their properties. 
The QCD sum rule analysis was developed by Shifman \textit{et al.} \cite{Shifman:1978bx,Shifman:1978by} 
and applied to the baryonic channel by Ioffe \cite{Ioffe:1981kw}. 
After that, the method was applied to the analyses in nuclear matter \cite{Drukarev:1988kd}. 
The $\Lambda_Q$ baryon (Q denotes the heavy quark in general) 
was first investigated by Shuryak \cite{Shuryak:1981fza} in the heavy quark mass limit. 
Since then, the sum rule was continuously improved \cite{Grozin:1992td,Bagan:1993ii} and 
the $1/m_Q$ corrections \cite{Dai:1995bc} and 
$\alpha_s$ corrections \cite{Groote:1996em} were taken into account. 
QCD sum rules with finite heavy quark mass were also constructed in Ref.\,\cite{Bagan:1992tp} and extended to 
the inclusion of the first order $\alpha_s$ corrections \cite{Groote:1999zp,Groote:2000py,Groote:2008dx} and parity projection \cite{Wang:2010fq}. 
Recently, the analyses are applied to the study of $\Lambda_c$ in nuclear matter \cite{Wang:2011hta,Azizi:2016dmr}. 
However, their results are not consistent with each other. The result of Ref.\,\cite{Wang:2011hta} indicates that 
the energy of the $\Lambda_c$ baryon increases and thus $\Lambda_c$ feels a net repulsive potential. 
On the other hand, the result of Ref.\,\cite{Azizi:2016dmr} shows that there is large  attractive interaction. 
For both cases, the $\alpha_s$ corrections, which are pointed out as large contributions \cite{Groote:1999zp,Groote:2000py,Groote:2008dx}, are not included and 
parity projection is not done. 
We carry out a new analysis of the parity projected $\Lambda_c$, $\Lambda$ and $\Lambda_b$ QCD sum rules including the $\alpha_s$ corrections in nuclear matter with the Gaussian kernel. 
Advantages of using the Gaussian sum rule were discussed in Refs.\,\cite{Bertlmann:1984ih,Orlandini:2000nv,Ohtani:2011zz}. 

The paper is organized as follows. 
In Sec.\ \ref{Sec: correlator}, we introduce the $\Lambda_c$ correlation function 
and construct the parity-projected in-medium QCD sum rules. 
The results of the analyses are summarized in Sec.\ \ref{Sec: result}, where the density dependence of the 
energy (the pole position of the $\Lambda_c$ propagator), the effective mass and the vector self-energy are presented. 
Next, we investigate effects of the density dependence of the four-quark condensate to the results in Sec.\ \ref{Sec: discussion}. 
We apply our analyses to $\Lambda$ in nuclear matter to discuss the validity of the estimation of the in-medium modification of  the four quark condensate. 
The in-medium modification of  $\Lambda_b$ is also studied in the same section.  
Summary and conclusions are given in Sec.\ \ref{Sec: Summary}.

\section{$\Lambda_c$ correlation function \label{Sec: correlator}}
We consider the two point correlation function of $\Lambda_c$ in nuclear matter: 
\begin{equation}
\begin{split}
\Pi ^{T}(q) =& i\int e^{iqx} d^{4}x \\
                & \times \langle \Psi_0 (\rho, u^{\mu}) | T[J_{\Lambda_c} (x)\overline{J}_{\Lambda_c}(0)] | \Psi_0 (\rho, u^{\mu}) \rangle , 
\label{eq: time ordered correlator}
\end{split}
\end{equation}
where $\Psi_0 (\rho, u^{\mu})$ is the ground state of nuclear matter,
which is characterized by its velocity $u^{\mu}$ and the nucleon density $\rho$. 
$J_{\Lambda_c}$ is an interpolating operator which has the same quantum numbers of the $\Lambda_c$ ground state.  
Without derivative, there are three independent interpolating operators \cite{Shuryak:1981fza}:  
\begin{equation}
\begin{split}
J_{\Lambda_c}^{1} (x) &= \epsilon ^{abc} (u^{Ta} (x) C  d^{b} (x)) \gamma_5 c^{c} (x), \\
J_{\Lambda_c}^{2} (x) &= \epsilon ^{abc} (u^{Ta} (x) C \gamma _5 d^{b} (x) ) c^{c} (x), \\
J_{\Lambda_c}^{3} (x) &= \epsilon ^{abc} (u^{Ta} (x) C \gamma _5 \gamma_ \mu d^{b} (x)) \gamma ^ \mu c^{c} (x). \\
\end{split}
\end{equation}
Here, $a$, $b$ and $c$ are color indices, $C = i\gamma_0 \gamma_2 $ stands for
the charge conjugation matrix and the spinor indices are omitted for simplicity. 
Each interpolating operator has a different type of di-quark, namely the $J_{\Lambda_c}^{1} (x)$, $J_{\Lambda_c}^{2} (x)$ and 
$J_{\Lambda_c}^{3} (x)$ contain the component of the pseudo scalar, the scalar and the vector di-quarks, respectively. 
Based on the feature that the scalar di-quark $\epsilon ^{abc} (q^{Ta}C \gamma _5 q^{b})$ is the most attractive channel 
\cite{Jaffe:2004ph,DeRujula:1975qlm,DeGrand:1975cf,tHooft:1976snw,Shuryak:1981ff}, one naturally expects 
that the $J_{\Lambda_c}^{2}$ strongly couples to the $\Lambda_c$ ground state. 
In fact, previous QCD sum rule studies of $\Lambda_c$ used the $J_{\Lambda_c}^{2}$ and the analyses of the mass in the vacuum were successful \cite{Wang:2010fq,Wang:2011hta,Zhang:2008iz}. 
A lattice QCD simulation \cite{Bali:2015lka} discusses the coupling between interpolating operators and hadron states and shows that the $J_{\Lambda_c}^{2}$ 
couples to the $\Lambda_c$ ground state. 
Therefore, we employ $J_{\Lambda_c}^{2}$ as the $\Lambda_c$ interpolating operator. 
In the following, we denote $J_{\Lambda_c}^{2}$ as $J_{\Lambda_c}$ for simplicity. 
With the help of the Lorentz covariance, parity invariance and time reversal invariance of the nuclear matter ground state, 
the correlation function of Eq.\,(\ref{eq: time ordered correlator}) can be decomposed into three scalar functions \cite{Cohen:1994wm}: 
 \begin{equation}
\begin{split}
\Pi ^{T}(q) =  \slash{q} \Pi^{T}_{1} (q_0, |\vec{q}|) + \Pi^{T}_{2} (q_0, |\vec{q}| )+ \slash{u} \Pi^{T}_{3} (q_0, |\vec{q}|)
\label{eq: time ordered}
\end{split}
\end{equation}
The variables of each scalar function, $\Pi_{i}^{T} \ (i=1, 2, 3)$, are $(q^2, q\cdot u)$, but we write them as $(q_0, |\vec{q}|)$ because 
we later take the rest frame of nuclear matter.  

Generally, a baryon interpolating operator couples both to positive parity states and negative parity states 
because the positive parity interpolating operator $J^{+} (x)$ is related to the negative parity interpolating operator: 
$J^{+} (x) = \gamma_5 J^{-} (x)$ \cite{Chung:1981cc}. 
The extra $\gamma_5$ only changes the sign of $\Pi^{T}_{2}$ in Eq.\,(\ref{eq: time ordered}). 
The method of the parity projection was proposed by Jido \textit{et al}. \cite{Jido:1996ia} and Kondo \textit{et al}. \cite{Kondo:2005ur} and 
was successful in investigating the mass of the nucleon ground state and its negative parity excited state in vacuum. 
The parity projected QCD sum rule has been improved to include the $\alpha_s$ corrections \cite{Ohtani:2012ps} and was applied to the analyses in nuclear matter \cite{Ohtani:2016pyk}. 

The parity projected QCD sum rule can be constructed from the ``forward-time'' correlation function 
\footnote{Note that in Ref.\,\cite{Jido:1996ia}, this correlator was called the ``old-fashioned'' correlator. } \cite{Jido:1996ia,Ohtani:2012ps,Ohtani:2016pyk}: 
\begin{equation}
\begin{split}
\Pi_m(q_0, |\vec{q}|) &= i \int d^{4}x e^{iqx} \theta(x_0) \\ 
                            & \hspace{0.5cm} \times \langle \Psi_0 (\rho, u^{\mu}) |T[\eta(x)\overline{\eta}(0)]| \Psi _0 (\rho, u^{\mu}) \rangle \\
                            = \slash{q} \Pi_{m1} &(q_0, |\vec{q}|) + \Pi_{m2} (q_0, |\vec{q}| )+ \slash{u} \Pi_{m3} (q_0, |\vec{q}|)
\end{split}
\label{eq:forward_m}
\end{equation}
The essential difference from the time-ordered correlation function is the insertion of
the Heaviside step-function $\theta (x_0)$ before carrying out the Fourier transform. 
This correlator contains contributions only from the states which propagate forward in time. 
Operating the parity projection operator $\mathrm{P}^{\pm}=\frac{\gamma_0 \pm1}{2}$ to $\Pi_m(q_0, |\vec{q}|)$ and taking the trace over the spinor index, 
the parity projected correlation functions can be derived: 
\begin{equation}
\begin{split}
\Pi _{m}^{+}(q_0, |\vec{q}|) &\equiv q_0 \Pi_{m1}(q_0, |\vec{q}|) +\Pi_{m2}(q_0, |\vec{q}|) \\
                                   &\hspace{2.58cm} + u_0 \Pi_{m3}(q_0, |\vec{q}|)\\
\Pi _{m}^{-}(q_0, |\vec{q}|) &\equiv q_0 \Pi_{m1}(q_0, |\vec{q}|) -\Pi_{m2}(q_0, |\vec{q}|) \\
                                   &\hspace{2.58cm} + u_0 \Pi_{m3}(q_0, |\vec{q}|). 
\end{split}
\label{eq:ppmfinal}
\end{equation}
Note that the parity projection can be carried out in accordance with that in vacuum 
because it is based on the invariance of the ground state of nuclear matter under the parity transformation.  

The QCD sum rule can be derived from the analyticity of the correlation function.  
The correlation functions $\Pi ^{\pm}_{m\mathrm{OPE}}$ which are calculated by the operator product expansion (OPE) in deep Euclidean region ($q^2 \rightarrow - \infty$) 
can be expressed by the hadronic spectral function $\rho^{\pm}_{m}$ by the use of the dispersion relation: 
\begin{equation}
\begin{split}
&\int _{-\infty}^{\infty} \mathrm{Im}[\Pi ^{\pm}_{m\mathrm{OPE}} (q_0, |\vec{q}|)]  W(q_0 ) dq_0 \\
&\hspace{3cm}= \pi \int _{0} ^{\infty} \rho^{\pm}_{m} (q_0, |\vec{q}|) W( q_0 ) dq_0. 
\label{Eq: dispersion relation}
\end{split}
\end{equation}
Here, we have introduced a weighting function $W( q_0 )$, which is real at real $q_0$ and 
analytic in the upper half of the imaginary plane of $q_0$. 
The details of the derivation of Eq.\,(\ref{Eq: dispersion relation}) are explained in Ref.\,\cite{Ohtani:2012ps}. 
Using the above equation, we investigate the spectral function and the in-medium properties of $\Lambda_c$. 
The specific forms of $\Pi ^{\pm}_{m\mathrm{OPE}}$ and $\rho^{\pm}_{m}$ will be discussed in the next susbsections. 
 
\subsection{OPE of the $\Lambda_c$ correlation function}
We first calculate the dimension 7 and 8 terms in the time-ordered correlation function 
and then construct the OPE of  the ``forward-time'' correlation function including the first order $\alpha_s$ corrections in nuclear matter taken 
from the previous studies of Refs.\,\cite{Wang:2011hta,Groote:2008dx}: 
\begin{equation}
\begin{split}
&\Pi _{m\mathrm{OPE}}(q_0, |\vec{q}|) \\
 &\hspace{1cm}=i\int e^{iqx}dx \theta (x_0)  \langle \Psi_0 |T\{J_{\Lambda_c}(x) \overline{J}_{\Lambda_c}(0)\} | \Psi_0 \rangle  \\
               &\hspace{1cm} = \slash{q}\Pi _{m1\mathrm{OPE}}(q_0, |\vec{q}|) + \Pi _{m2\mathrm{OPE}}(q_0,|\vec{q}|) \\ 
              & \hspace{1.4cm}+ \slash{u}\Pi _{m3\mathrm{OPE}}(q_0,|\vec{q}|) \\
\label{eq:nemui}
\end{split}
\end{equation}
Explicit expressions of $\rho_{mi\mathrm{OPE}} \equiv \frac{1}{\pi} \mathrm{Im} [\Pi_{mi\mathrm{OPE}}] \ (i=1, 2, 3)$ are given as 
\begin{equation}
\begin{split}
&q_0 {\rho_{m1\mathrm{OPE}} (q_0, |\vec{q}|)}|_{\vec{q}=0} \\
&\hspace{1cm} = q_0 \rho_{m1\mathrm{OPE}}^{pert}(q_0) + q_0 \rho_{m1\mathrm{OPE}}^{cond}(q_0), 
\end{split}
\end{equation}
\begin{equation}
\begin{split}
q_0 \rho_{m1\mathrm{OPE}}^{pert}(q_0) =& \frac{q_0^5}{128\pi^4}\Big\{\rho ^0_{m1\mathrm{OPE}} (\frac{m_c^2}{q_0^2}) \left( 1+\frac{\alpha_s}{\pi}\ln\frac{\mu^2}{m_c^2}\right) \\
                                                       &+\frac{\alpha_s}{\pi}\rho ^1_{m1\mathrm{OPE}} (\frac{m_c^2}{q_0^2})\Big\} \theta(q_0-m_c), 
\end{split}
\end{equation}
\begin{equation}\label{lead0q}
\rho ^0_{m1\mathrm{OPE}} (z)=\frac14-2z+2z^3-\frac14z^4-3z^2\ln z, 
\end{equation}
\begin{equation}
\begin{split}
&\rho ^1_{m1\mathrm{OPE}} (z) = \frac{71}{48}-\frac{565}{36}z-\frac{7}{8}z^2+\frac{625}{36}z^3
-\frac{109}{48}z^4 \\
 &  \ \ - \left(\frac{49}{36}-\frac{116}{9}z+\frac{116}{9}z^3-\frac{49}{36}z^4
  \right)\ln(1-z) \\ 
  & \ \ +\left(\frac{1}{4}-\frac{17}{3}z-11z^2+\frac{113}{9}z^3-\frac{49}{36}z^4\right)\ln z \\
  & \ \ +\frac{2}{3}\left(1-8z+8z^3-z^4\right)\left(\mathrm{Li}_2(z)+\frac12\ln(1-z)\ln z\right) \\
  & \ \ -\frac{1}{3}z^2\left(54+8z-z^2\right)\left(\mathrm{Li}_2(z)-\zeta(2)+\frac{1}{2}\ln^2z\right) \\
  & \ \ -12z^2\left(\mathrm{Li}_3(z)-\zeta(3)-\frac{1}{3}\mathrm{Li}_2(z)\ln(z)\right), 
\end{split}
\end{equation}
\begin{equation}
\begin{split}
& q_0 \rho_{m1\mathrm{OPE}}^{cond}(q_0) = 
      -\frac{m_c^2}{768\pi^2}\langle \frac{\alpha_sGG}{\pi}\rangle_{m} \\
& \hspace{2.5cm} \times \int_{\frac{m_c^2}{q_0^2}}^1 d\alpha \frac{(1-\alpha)^2}{\alpha^2} \delta (q_0- \frac{m_c}{\sqrt{\alpha}}) \\
                                    &\ \ +\frac{q_0}{128\pi^2}\langle \frac{\alpha_sGG}{\pi}\rangle_{m}  \int_{\frac{m_c^2}{q_0^2}}^1 d\alpha \alpha \theta (q_0 -m_c) \\
                                    &\ \ -q_0 \frac{\langle q^{\dagger}iD_0q\rangle_{m} }{6\pi^2} \int_{\frac{m_c^2}{q_0^2}}^1 \alpha (1-\alpha ) d\alpha \theta (q_0 -m_c) \\
                                    &\ \ +\frac{\langle q^{\dagger}iD_0q\rangle_{m} }{6\pi^2} \int_{0}^1 d\alpha \alpha (1-\alpha) \frac{3}{2} q_0^2  \delta (q_0- \frac{m_c}{\sqrt{\alpha}}) \\
                                    &\ \ +\frac{ \langle\bar{q}q\rangle_{m} ^2 + \langle q^\dagger q\rangle_{m} ^2 }{12}\delta(q_0-m_c) \\
                                    &\ \ + \frac{1}{2^3 \cdot 3} \langle \overline{q}g \sigma G q \rangle _{m}  \langle \overline q q \rangle_{m}  \\
                        & \ \ \times \Bigg ( \frac{1}{8} \Big ( \delta ^{''} (q_0 - m_c) - \frac{7}{m_c} \delta ^{'} (q_0 - m_c) \\
                        &\hspace{4cm} + \frac{6}{m_c^2} \delta (q_0 - m_c)\Big ) \Bigg ) \\
                                    &\ \  -q_0 \Bigg [  \frac{q_0 \langle q^\dagger q\rangle _{m} }{4\pi^2} \int_{\frac{m_c^2}{q_0^2}}^1 d\alpha \alpha (1-\alpha) \theta (q_0 -m_c) \\
                                    &\ \ +\frac{1}{8\pi^2} \left (\langle q^\dagger iD_0iD_0 q\rangle_{m} +\frac{1}{12} \langle q^{\dagger} g_s\sigma G q\rangle_{m} \right ) \\ 
                                    &\ \  \times  \int_{0}^1 \alpha (1- \alpha ) \left ( \delta (q_0- \frac{m_c}{\sqrt{\alpha}} ) -\frac{\alpha q_0^3}{m_c^{2}} \delta^{'} (q_0 - \frac{m_c}{\sqrt{\alpha}} ) 
                                       \right ) d\alpha \\
                                    &\ \ -\frac{1}{96\pi^2} \langle q^{\dagger} g_s\sigma G q\rangle_{m} \int_{0}^1 \alpha (1-\alpha) \\
                                    &\ \  \times \left ( 9 \delta (q_0 - \frac{m_c}{\sqrt{\alpha}} ) + \frac{\alpha q_0^3}{m_c^{2}} \delta^{'} (q_0 - \frac{m_c}{\sqrt{\alpha}} ) \right ) d\alpha \\
                                    &\ \  +\frac{\langle q^{\dagger } g_s\sigma G q\rangle_{m} }{32\pi^2} \int_{0}^1 \alpha \delta (q_0- \frac{m_c}{\sqrt{\alpha}} ) d\alpha \Bigg ]
\end{split}
\end{equation}
and 
\begin{equation}
\rho _{m2\mathrm{OPE}}(q_0, |\vec{q}|)|_{\vec{q}=0} = \rho _{m2\mathrm{OPE}}^{pert}(q_0) + \rho_{m2\mathrm{OPE}}^{cond}(q_0), 
\end{equation}
\begin{equation}
\begin{split}
\rho_{m2\mathrm{OPE}}^{pert} (q_0) =& \frac{m_c q_0^4}{128\pi^4}\Big\{\rho ^{0} _{m2\mathrm{OPE}} (\frac{m_c^2}{q_0^2}) \left( 1+\frac{\alpha_s}{\pi}\ln\frac{\mu^2}{m_c^2}\right) \\
  &+\frac{\alpha_s}{\pi}\rho ^{1}_{m2\mathrm{OPE}} (\frac{m_c^2}{q_0^2})\Big\} \theta(q_0-m_c), 
\end{split}
\end{equation}
\begin{equation}\label{lead0m}
\rho ^0_{m2\mathrm{OPE}}(z)=1+9z-9z^2-z^3+6z(1+z)\ln z,  
\end{equation}
\begin{equation}
\begin{split}
&\rho^1_{m2\mathrm{OPE}}(z) = 9+\frac{665}{9}z-\frac{665}{9}z^2-9z^3 \\
  & \ \ -\left(\frac{58}{9}+42z-42z^2-\frac{58}{9}z^3\right)\ln(1-z) \\
  & \ \ +\left( 2+\frac{154}{3}z-\frac{22}{3}z^2-\frac{58}{9}z^3 \right)\ln z \\
  &+\frac{8}{3}\left(1+9z-9z^2-z^3\right)\left(\mathrm{Li}_2(z)
  +\frac{1}{2}\ln(1-z)\ln z\right) \\ 
  & \ \ +z\left(24+36z+\frac{4}{3}z^2\right) \left(\mathrm{Li}_2(z)-\zeta(2)+\frac{1}{2}\ln^2z\right) \\
  & \ \  +24z(1+z)\left(\mathrm{Li}_3(z)-\zeta(3)-\frac{1}{3}\mathrm{Li}_2(z)\ln z\right), 
\end{split}
\end{equation}
\begin{equation}
\begin{split}
&\rho_{m2\mathrm{OPE}}^{cond}(q_0)= -\frac{m_c}{768\pi^2}\langle \frac{\alpha_sGG}{\pi}\rangle _{m} \\
                                    & \hspace{2.5cm} \times \int_{\frac{m_c^2}{q_0^2}}^1 d\alpha \frac{(1-\alpha)^2}{\alpha}\frac{m_c}{\sqrt{\alpha}} \delta(q_0-\frac{m_c}{\sqrt{\alpha}}) \\
                                    &\ \ +\frac{m_c}{192\pi^2}\langle \frac{\alpha_sGG}{\pi}\rangle _{m}  \int_{\frac{m_c^2}{q_0^2}}^1 d\alpha \frac{(1-\alpha )^3}{\alpha ^2} \theta (q_0 -m_c) \\ 
                                    &\ \ +\frac{m_c}{128\pi^2}\langle \frac{\alpha_sGG}{\pi}\rangle _{m}  \int_{\frac{m_c^2}{q_0^2}}^1 d\alpha \theta (q_0 -m_c) \\
                                    &\ \ +\frac{m_c \langle q^{\dagger}iD_0q\rangle_{m}}{4\pi^2} \int_{0}^1(1-\alpha )  q_0 \delta (q_0- \frac{m_c}{\sqrt{\alpha}}) d\alpha \\
                                    &\ \ +\frac{ \langle\bar{q}q\rangle_{m} ^2 + \langle q^\dagger q\rangle_{m} ^2 }{12}\delta(q_0-m_c) \\
                                    &\ \ + \frac{1}{2^3 \cdot 3} \langle \overline{q}g \sigma G q \rangle_{m}  \langle \overline q q \rangle_{m}  \\ 
                                    &\ \ \times \Bigg ( 
                                        \frac{1}{8} \Big (\delta^{''} (q_0 - m_c) -\frac{3}{m_c} \delta^{'} (q_0 - m_c) \\
                                    & \hspace{4cm} + \frac{3}{m_c^2} \delta (q_0 - m_c) \Big ) \Bigg ) \\
                                    &\ \ -q_0 \Bigg [ m_c \frac{ \langle q^\dagger q\rangle_{m}}{4\pi^2} \int_{\frac{m_c^2}{q_0^2}}^1 (1-\alpha )d\alpha \theta (q_0 -m_c) \\
                                    &\ \ +\frac{1}{4\pi^2} \left(\langle q^\dagger iD_0iD_0 q\rangle_{m} +\frac{1}{12} \langle q^{\dagger} g_s\sigma G q\rangle_{m} \right) \\
                                    &\ \ \times \int_{0}^1  2 (1-\alpha) 
                                       \Big ( \frac{\alpha q_0^2}{4m_c} ( \frac{\sqrt{\alpha}}{m_c} \delta (q_0 - \frac{m_c}{\sqrt{\alpha}} ) \\ 
                                        & \hspace{3.8cm} + \delta^{'} (q_0 - \frac{m_c}{\sqrt{\alpha}} ) ) \Big ) d\alpha \\
                                    &\ \  -\frac{\langle q^{\dagger } g_s\sigma G q\rangle_{m} }{96\pi^2} \int _{0}^1 \Big ( 7 \sqrt{\alpha} \delta (q_0- \frac{m_c}{\sqrt{\alpha}}) \\
                                    & \hspace{3.2cm}    +\frac{\alpha q_0^2}{m_c} \delta^{'} (q_0- \frac{m_c}{\sqrt{\alpha}}) \Big )d\alpha \\
                                     &\ \ +\frac{\langle q^{\dagger } g_s\sigma G q\rangle _{m} }{32\pi^2} \int _{0}^1  \sqrt{\alpha} \delta (q_0- \frac{m_c}{\sqrt{\alpha}} )d\alpha 
\Bigg ]
\end{split}
\end{equation}
and 
\begin{equation}
\begin{split}
&{\rho_{m3\mathrm{OPE}}(q_0, |\vec{q}|)}|_{\vec{q}=0}  = -\frac{\langle q^{\dagger} q\rangle_{m}}{8\pi^2} \\
                                                      &\hspace{1cm} \times \int_{\frac{m_c^2}{q_0^2}}^1 \alpha (1-\alpha ) (q_0^2- \frac{m_c^2}{\alpha}) d\alpha \theta (q_0 -m_c) \\ 
                                                       &\ \ -\frac{5}{8\pi^2} \left(\langle q^\dagger iD_0iD_0 q\rangle_{m} +\frac{1}{12} \langle q^{\dagger} g_s\sigma G q\rangle_{m} \right) \\
                                                       &\ \ \times \int_{0}^1 \alpha (1-\alpha ) q_0\delta (q_0 - \frac{m_c}{\sqrt{\alpha}}) d\alpha \\
                                                       &\ \ +\frac{\langle q^{\dagger } g_s\sigma G q\rangle _{m} }{96\pi^2} \int_{0}^1 \alpha (1- \alpha ) q_0 \delta (q_0 - \frac{m_c}{\sqrt{\alpha}}) d\alpha \\
                                                       &\ \ -\frac{\langle q^{\dagger } g_s\sigma G q\rangle _{m} }{32\pi^2} \int_{\frac{m_c^2}{q_0^2}}^1 \alpha d\alpha \theta (q_0 -m_c) \\
                                                       &\ \ + q_0 \ \frac{2\langle q^{\dagger}iD_0 q\rangle_{m}}{3\pi^2} \int_{\frac{m_c^2}{q_0^2}}^1 \alpha (1-\alpha ) d\alpha \theta (q_0 -m_c), 
\end{split}
\end{equation}
where $\mathrm{Li}_{s}(z)$ is Polylogarithm and $\zeta(s)$ is Riemann's zeta function. 
The matrix elements $\langle \mathcal{O} \rangle _m$ stand for the 
expectation values of the operators $\mathcal{O}$ in nuclear matter. 
We have used the factorization hypothesis for the four-quark and quark-gluon mixed operators 
in these equations and will discuss its validity in section \ref{Sec: discussion}. 
The dimension 7 condensate $\langle \frac{\alpha_sGG}{\pi}\rangle _{m} \langle \overline{q} q \rangle _{m}$ 
term does not appear because its Wilson coefficient is equal to zero. 
The Wilson coefficients of $\langle q^{\dagger} q\rangle_{m}$, $\langle q^{\dagger}iD_0q\rangle_{m}$, $\langle q^{\dagger}iD_0iD_0q\rangle_{m}$, 
$\langle q^{\dagger}g_s\sigma G q\rangle_{m}$ terms are different from those in the literature \cite{Wang:2011hta}. 
However, our results are consistent with the OPE of $\Lambda$ \cite{Jeong:2016qlk} 
in the limit $m_c^2 \rightarrow 0$. 
Note that, in the case of $\Lambda$, extra contributions of $\langle \overline{s}s \rangle _{m}$ condensate appear and 
the Wilson coefficients of the gluon condensate are modified. 

The values of the parameters in OPE are summarized below.
In the linear density approximation, which is valid at sufficiently low density \cite{Cohen:1991nk,Drukarev:1988kd}, 
the expectation values of the condensates in nuclear matter $\langle \mathcal{O} \rangle _m$ 
are given as  
\begin{equation}
\begin{split}
\langle \overline{q}q\rangle _{m} &= \langle \overline{q}q\rangle _{0} + \rho \langle \overline{q}q\rangle _{N} =\langle \overline{q}q\rangle _{0} + \rho \frac{\sigma_ N}{2m_q} \\
\langle \overline{s}s\rangle _{m} &= \langle \overline{s}s\rangle _{0} + \rho \langle \overline{s}s\rangle _{N} \\
\langle q^{\dagger} q\rangle _{m} &= \rho \frac{3}{2} \\
\langle \frac{\alpha _s}{\pi }G^{2}\rangle _{m} &= \langle \frac{\alpha _s}{\pi }G^{2}\rangle _{0} + \rho \langle \frac{\alpha _s}{\pi }G^{2}\rangle _{N}  \\
\langle q^\dagger iD_0q\rangle _{m} &= \rho \langle q^\dagger iD_0q\rangle _{N}  =\rho \frac{3}{8} M_N A_2^q \\
\langle \overline{q} i D_0 q\rangle _{m} &=  m_q \langle q^{\dagger} q \rangle _{m}  \simeq 0\\
                                                                       \langle q^\dagger g\sigma \cdot Gq\rangle _{m} &=  \rho \langle q^\dagger g\sigma \cdot Gq\rangle _{N} \\
\langle q^\dagger i D_0 i D_0 q\rangle  _{m} +&\frac{1}{12}\langle q^\dagger g\sigma \cdot Gq\rangle _{m} \\
=  \bigl ( &\langle q^\dagger i D_0 i D_0 q\rangle _N +\frac{1}{12}\langle q^\dagger g\sigma \cdot Gq\rangle _N \bigr ) \rho \\
=  \rho & \frac{1}{4} M_N^2 A_3^q 
\end{split}
\label{eq:condensates1}
\end{equation}
where $\langle \mathcal{O} \rangle_0$ and $\langle \mathcal{O} \rangle_N$ are, respectively, the 
vacuum and nucleon expectation values of the operator $\mathcal{O}$, and $\langle \overline{q} \cdots q \rangle$ and $\langle q^{\dagger} \cdots q \rangle$ 
denotes the average over the up and down quarks, $\frac{1}{2} \left (\langle \overline{u} \cdots u \rangle + \langle \overline{d} \cdots d \rangle \right )$ and 
$\frac{1}{2} \left (\langle u^{\dagger} \cdots u \rangle + \langle d^{\dagger} \cdots d \rangle \right )$, respectively. 
The quantities $A_2^q$ and $A_3^q$ can be expressed as 
moments of the parton distribution functions \cite{Cohen:1994wm}. 
For the quark condensate $\langle \overline{q} q \rangle$, higher order density terms have been calculated 
by chiral perturbation theory \cite{Kaiser:2007nv,Goda:2013bka}. 
However, their contributions are small up to the normal nuclear mater density and thus 
we do not take them into account in this study. 
The numerical values of the parameters appearing in Eq.\,(\ref{eq:condensates1}) are given in Table\,\ref{tab:the values of the condensate in nuclear matter}. 
\begin{table}[t!]
\begin{center}
\begin{tabular}{|c|c|}
\hline 
parameters & values \\ \hline
$\langle \overline{q}q\rangle _{0}$  & $-(0.246\pm 0.002 \mathrm{GeV})^3$ \cite{Borsanyi:2012zv} \\ \hline
$m_q$  & $4.725\mathrm{MeV}$ \cite{Agashe:2014kda}  \\ \hline
$\sigma _N$  & $45 \mathrm{MeV}$   \\ \hline
$\langle q^{\dagger} q\rangle _{m}$ &  $\rho \frac{3}{2}$ \\ \hline
$\langle \frac{\alpha _s}{\pi }G^{2}\rangle _{0}$ & $0.012 \pm 0.0036 \mathrm{GeV}^4$  \cite{Colangelo:2000dp} \\ \hline 
$\langle \frac{\alpha _s}{\pi }G^{2}\rangle _{N}$ & $-0.65 \pm 0.15\mathrm{GeV}$ \cite{Jin:1993up} \\ \hline 
$A^q_2$ & $0.62 \pm 0.06$ \cite{Martin:2009iq} \\ \hline
$A^q_3$ & $0.15 \pm 0.02$ \cite{Martin:2009iq} \\ \hline
$\langle q^\dagger g\sigma \cdot Gq\rangle _{N}$ & $-0.33 \mathrm{GeV}^2$ \cite{Jin:1993up} \\ \hline
$m_c$& $1.67\pm 0.07 \mathrm{GeV}$ \cite{Agashe:2014kda} \\ \hline
$\alpha_s$& $0.5 $ \\
\hline
\end{tabular}
\caption{Values of parameters appearing in Eq.\,(\ref{eq:condensates1}). }
\label{tab:the values of the condensate in nuclear matter}
\end{center}
\end{table}

\subsection{Phenomenological side of the $\Lambda_c$ correlation function}
The correlation function at the physical energy region ($q^2 > 0$ ) can be described by the hadronic degrees of freedom. 
In this study, we use the so called ``pole + continuum'' ansatz for the correlator in such energy region. 
The pole stands for the contributions of the ground state and is assumed to be proportional to 
the single $\Lambda_c$ baryon propagator $G(q)$ in vacuum, 
\begin{equation}
\begin{split}
G(q) = \frac{\slash q + M_{\Lambda_c}}{(q^2 - M_{\Lambda_c}^2 +i \epsilon )}. 
\end{split}
\end{equation}
Then the pole contribution in the spectral function $\rho^{\mathrm{T}}(q) \equiv \frac{1}{\pi} \mathrm{Im} [\Pi ^{\mathrm{T}}(q)]$ is 
\begin{equation}
\begin{split}
- \frac{1}{\pi} \mathrm{Im} &\left [ |\lambda^2|  \frac{\slash q + M_{\Lambda_c}}{(q^2 - M_{\Lambda_c}^2 +i \epsilon )}  \right ] \\
& \ \ =  |\lambda^2| (\slash q + M_{\Lambda_c}) \delta (q^2 - M_{\Lambda_c}^2). 
\end{split}
\end{equation}
Here, $|\lambda^2|$ is the residue and gives the coupling strength between the ground state and the employed interpolating operator. 
Taking into account the contributions of the lowest lying negative parity state $\Lambda_c^{-}$, the phenomenological side of the spectral function 
in the ``pole + continuum'' ansatz can be expressed as 
\begin{equation}
\begin{split}
\rho^{T} (q) = & |\lambda^2_{+}| (\slash q + M_{\Lambda_c}) \delta (q^2 - M_{\Lambda_c}^2) \\
                   &+ |\lambda^2_{-}| (\slash q - M_{\Lambda_c^{-}}) \delta (q^2 - M_{\Lambda_c^{-}}^{2}) \\
                   &+ \frac{1}{\pi} \mathrm{Im} \left [ \Pi _{\mathrm{OPE}} (q) \right ] \theta (q^2 - q_{th}^2 ). 
\end{split}
\end{equation}
For the continuum states, the quark hadron duality is assumed. 

In nuclear matter, the $\Lambda_c$ propagator can be described as 
\begin{equation}
\begin{split}
G(q_0, |\vec{q}|) = \frac{Z^{'}(q_0, |\vec{q}|)}{\slash q - M - \Sigma (q_0, |\vec{q}|) + i\epsilon} 
\label{Eq:npropagator}
\end{split}
\end{equation}
where $\Sigma(q_0, |\vec{q}|)$ is the self-energy and $Z^{'}(q_0, |\vec{q}|)$ denotes the renormalization factor of 
the wave function. 
As in Eq.\,(\ref{eq:forward_m}), the self-energy can be decomposed as 
\begin{equation}
\begin{split}
\Sigma (q_0, |\vec{q}|) =& \Sigma ^{s'} (q_{0}, |\vec{q}|) + \Sigma ^{v'} (q_{0}, |\vec{q}|) \slash{u} \\
                                &+ \Sigma ^{q'} (q_{0}, |\vec{q}|) \slash{q}. 
\label{Eq:SRself}
\end{split}
\end{equation}
Renormalizing the $\Sigma ^{q'} (q_{0}, |\vec{q}|)$ contributions to the $Z^{'}(q_0, |\vec{q}|)$, $\Sigma ^{v'} (q_{0}, |\vec{q}|)$  and effective mass $M^{*}_{\Lambda_c}$, 
the in-medium $\Lambda_c$ baryon propagator $G(q_0, |\vec{q}|) $ is expressed as 
\begin{equation}
\begin{split}
&G(q_0, |\vec{q}|) \\
& \hspace{1cm} = Z(q_0, |\vec{q}|)\frac{\slash q - \slash u \Sigma_v + M^*_{\Lambda_c}}{(q_0 - E_{\Lambda_c} +i \epsilon )(q_0 + \overline{E}_{\Lambda_c} - i \epsilon )}, 
\label{eq: in-medium propagator}
\end{split}
\end{equation}
where
\begin{equation}
\begin{split}
E_{\Lambda_c} =& \Sigma_v + \sqrt{M^{*2}_{\Lambda_c} + \vec{q}^2}, \\
 \overline{E}_{\Lambda_c} =& -\Sigma_v + \sqrt{M^{*2}_{\Lambda_c} + \vec{q}^2}.
\label{eq: eenrgy}
\end{split}
\end{equation}

The contribution of the positive energy state in $G(q_0, |\vec{q}|) $ is
\begin{equation}
\begin{split}
Z(q_0, |\vec{q}|)\frac{1}{2\sqrt{M^{*2}_{\Lambda_c}+\vec{q}^2}}\frac{\gamma_0 E_{\Lambda_c} - \slash u \Sigma_v + M^*_{\Lambda_c}}{(q_0 - E_{\Lambda_c} +i \epsilon )}. 
\end{split}
\end{equation}
After taking the rest frame of nuclear matter, the forward-time spectral function in nuclear matter can be described as 
\begin{equation}
\begin{split}
& \rho (q_0, |\vec{q}|=0)  \\
&=\frac{ |\lambda^2_{+}|}{2 M^{*}_{\Lambda_c}} ( \gamma_0 E_{\Lambda_c} - \slash u \Sigma_v + M^*_{\Lambda_c}) \delta (q_0 - E_{\Lambda_c}) \\
                             & \ \ \ + \frac{ |\lambda^2_{-}|}{2 M^{*}_{\Lambda_c ^{-}}} ( \gamma_0 E_{\Lambda_c ^{-}} - \slash u \Sigma_v - M^{*}_{\Lambda_c^{-}}) \delta (q_0 - E_{\Lambda_c^{-}}) \\
                                  & \ \ \ + \frac{1}{\pi} \mathrm{Im} \left [ \Pi _{\mathrm{OPE}} (q_0, |\vec{q}|=0 ) \right ] \theta (q_0 - q_{th} ).
\label{Eq: Forward SPF}
\end{split}
\end{equation}

\subsection{Equation of $\Lambda_c$ QCD sum rule with parity projection}
As we have introduced in Eq.\,(\ref{Eq: dispersion relation}), 
the OPE description of Eq.\,(\ref{eq:nemui}) and the phenomenological description of Eq.\,(\ref{Eq: Forward SPF}) can be connected
with the help of the analyticity of the correlation function in $q_0$ plane. 
For specifying the kernel $W(q_0)$, we use the Gaussian sum rule and its equation is given as 
\begin{equation}
\begin{split}
G_{mOPE} ^{\pm} (\tau ) = G_{\mathrm{SPF}}^{\pm}(\tau), 
\label{eq: QCD sum rule}
\end{split}
\end{equation}
where 
\begin{equation}
\begin{split}
G_{\mathrm{SPF}}^{\pm}(\tau) &= \int _{0}^{\infty } \rho ^{\pm} (q_0) \frac{1}{\sqrt{4 \pi \tau}} \exp \left (- \frac{(q_0^2 - m_c^2)^2}{4 \tau} \right ) dq_0 
\end{split}
\end{equation}
and
\begin{equation}
\begin{split}
G_{mOPE} ^{\pm}(\tau ) \equiv G_{m1\mathrm{OPE}} (\tau ) \pm G_{m2\mathrm{OPE}} (\tau ) + G_{m3\mathrm{OPE}} (\tau )
\end{split}
\end{equation}
with 
\begin{equation}
\begin{split}
G_{m1\mathrm{OPE}} (\tau ) &= \int_{0}^{\infty} q_0 \rho_{m1\mathrm{OPE}} (q_0, |\vec{q}|) \\
& \hspace{1.2cm} \times \frac{1}{\sqrt{4\pi \tau}} \exp \left ( -\frac{(q_0^2 - m_c^2)^2}{4 \tau} \right ) dq_0 \\
G_{m2\mathrm{OPE}} (\tau ) &= \int_{0}^{\infty}  \rho_{m2\mathrm{OPE}} (q_0, |\vec{q}|) \\ 
& \hspace{1.2cm} \times \frac{1}{\sqrt{4\pi \tau}} \exp \left ( -\frac{(q_0^2 - m_c^2)^2}{4 \tau} \right ) dq_0 \\
G_{m3\mathrm{OPE}} (\tau ) &= \int_{0}^{\infty} \rho_{m3\mathrm{OPE}} (q_0, |\vec{q}|) \\
& \hspace{1.2cm} \times \frac{1}{\sqrt{4\pi \tau}} \exp \left ( -\frac{(q_0^2 - m_c^2)^2}{4 \tau} \right ) dq_0.
\end{split}
\end{equation}
Here $\rho ^{\pm} (q_0)$ is the hadronic spectral function and $\pm$ stands for the parity of the hadronic states, 
\begin{equation}
\begin{split}
\rho ^{\pm} (q_0) =& |\lambda ^{\pm}|^2 \delta (q_0-E_{\Lambda_c^{\pm}} ) \\ 
                         &+ \frac{1}{\pi} \mathrm{Im} \left [\Pi_{mOPE}^{\pm} (q_0) \right ] \theta (q_0 -q_{th}^{\pm}). 
\label{eq: pnspf}
\end{split}
\end{equation}

Generally, the kernel of the Gaussian sum rule is $\frac{1}{\sqrt{4\pi \tau}} \exp \left ( -\frac{(q_0^2 - s)^2}{4 \tau} \right )$ and contains two parameters, $\tau$ and $s$. 
We set the value of $s$ to $m_c^2$, and as a result, the exponential suppression starts from $m_c$. 
This is the different point from the case of the Borel sum rule where the kernel, $ \exp \left ( -\frac{q_0^2 }{M^2} \right )$, is used. 
One naively expects that the Gaussian sum rule strongly reflects on the behavior of the spectral function above $m_c$ 
more than the Borel sum rule. 
It is one advantage of using the Gaussian sum rule.  

We extract the values of parameters in $\rho ^{\pm} (q_0)$ from Eq.\,(\ref{eq: QCD sum rule}) 
by minimizing $\chi ^2$, defined as 
\begin{equation}
\chi ^2 =  \frac{1}{n_ {set}\times n_ \tau}\sum_{j=1} ^{n_ {set}}\sum_{i=1} ^{n_\tau}\frac{(G_{mOPE}^{j}(\tau _i) - G_{\mathrm{SPF}}^{j} (\tau _i ))^2 }{\sigma ^{j} (\tau _i)^2},
\end{equation}
where the errors of $G_{m\mathrm{OPE}}$, $\sigma(\tau_i)$, are evaluated based on the method proposed in Ref.\,\cite{Leinweber:1995fn}: 
\begin{equation}
\sigma^{j} (\tau _i)^2 = \frac{1}{n_ {set}-1}\sum_{j=1} ^{n_ {set}} (G_{mOPE}^{j}(\tau _i) - \overline{G}_{mOPE}(\tau _i))^2,
\end{equation}
with
\begin{equation}
\overline{G}_{mOPE}(\tau _i) = \frac{1}{n_{set}}\sum_{j=1} ^{n_ {set}} G_{mOPE}^{j}(\tau _i). 
\end{equation}
Here  $n_\tau$ and $n_{set}$ are the numbers of the point $\tau $ in the analyzed $\tau$ region and 
of the condensate sets which are randomly generated with errors, respectively. 

The parameter region of $\tau$ is chosen as follows. 
The lowest value of $\tau$ is determined by the convergence of the OPE while 
the highest value of $\tau$ is constrained to satisfy the pole dominance condition. 
These conditions will justify the truncation of the OPE. 
Specifically, we use the well established criterion that the ratio of the
highest dimensional term to the whole $G_{mOPE}$ is less than 0.1 and 
the ratio of the pole contribution to the whole $G_{mSPF}$ is more than 0.5.
The region $1.25 <\tau [\mathrm{GeV}^4]< 3.5 $ satisfies the above conditions in vacuum. 
We use the same parameter region for the analyses in nuclear matter. 

\section{Results of $\Lambda_c$ \label{Sec: result}}
\subsection{Behavior of OPE}
\begin{figure}[!tbp]
\begin{center}
  \includegraphics[width=9.cm]{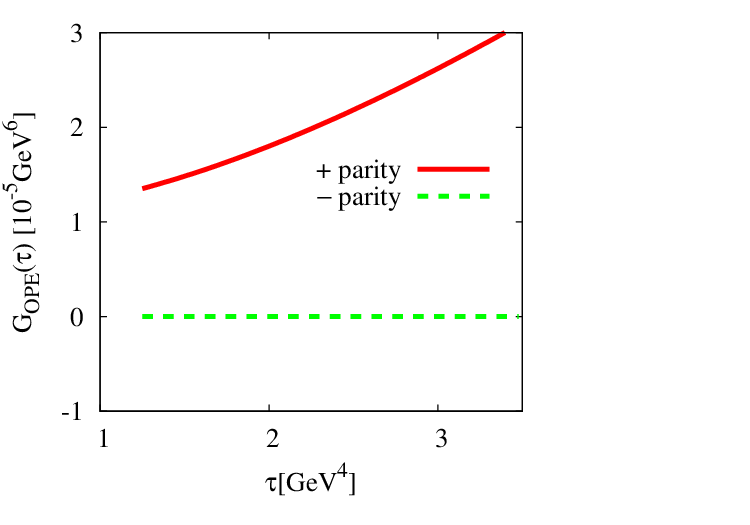}
 \end{center}
 \caption{The positive and negative parity OPE, $G^{\pm}_{m\mathrm{OPE}} (\tau)$, in vacuum. 
}
 \label{fig:OPE p and n}
\end{figure}
We first discuss the behaviors of $G_{m\mathrm{OPE}}^{\pm}(\tau)$ in vacuum, shown in Fig.\,\ref{fig:OPE p and n}. 
We find that the OPE of the positive parity correlation function is much larger than that of the negative parity state. 
In the hadronic degrees of freedom, this means that the interpolating operator $J_{\Lambda_c}$ 
couples mainly to the positive parity ground states, and 
that the structure of $\Lambda_c$ is similar to the structure of the interpolating operator. 
This result is consistent with the quark model picture where the angular momentum between the quarks are all S-wave. 
On the other hand, the present interpolating operator hardly couples to the negative parity states. 

\begin{figure*}[!tbp]
\begin{center}
  \includegraphics[width=13.cm]{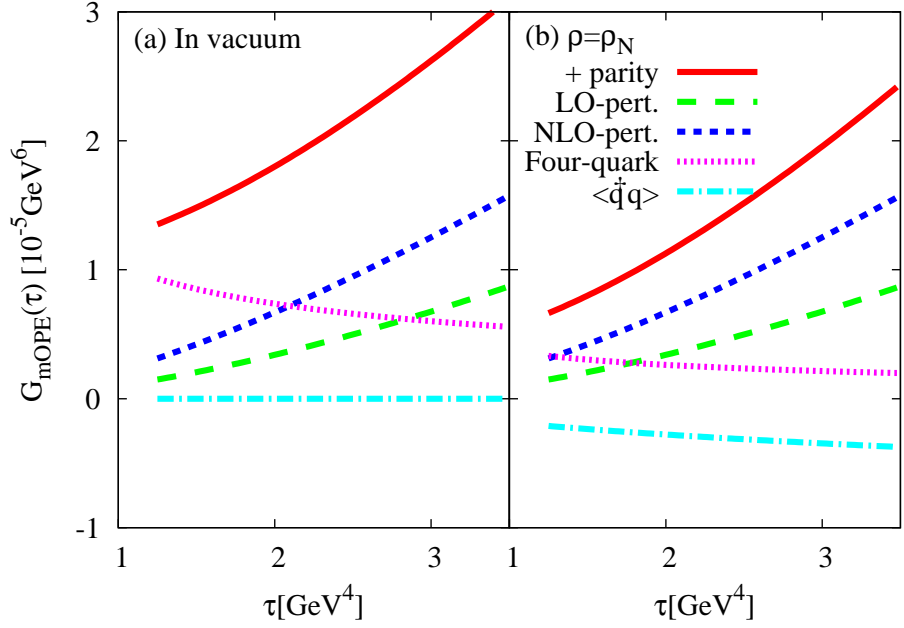}
 \end{center}
 \caption{The density dependences of the positive parity OPE $G^{+}_{m\mathrm{OPE}} (\tau)$. 
The LO-perturbative, the NLO-perturbative, the four-quark and the vector quark condensate $\langle q^{\dagger }q\rangle$ terms are 
shown. Here, $\rho_N$ means normal nuclear matter density. 
}
 \label{fig: OPE density charm}
\end{figure*}

$G_{m\mathrm{OPE}}^{+}(\tau)$ in vacuum and at normal nuclear matter density are shown in 
Fig.\,\ref{fig: OPE density charm}.  
The LO-perturbative, the NLO-perturbative, the four-quark and the vector quark condensate $\langle q^{\dagger }q\rangle$ terms 
whose contributions are large, are also shown in the same figure. 
We find that the $\alpha_s$ correction on the pertrubative terms are as large as twice of the leading order term. 
As for higher order $\alpha_s$ corrections, 
the large $\alpha_s$ corrections in the perturbative term affect the behavior of continuum states. 
In conventional QCD sum rule analyses, the continuum states are separated from the lowest lying state by the threshold parameter. 
Therefore, adjusting the value of the threshold parameter is expected to compensate for the effects from higher order $\alpha_s$ corrections. 
Among the non-perturbative contributions, 
the four-quark condensate is dominant in vacuum. 
Fig.\,\ref{fig: OPE density charm} shows that the 
in-medium modification of the four-quark condensate and  $\langle q^{\dagger }q\rangle_m$ is the main origin of the density dependence of  $G^{+}_{m\mathrm{OPE}} (\tau)$. 
Therefore, we find that 
the charm quark plays less role than the light quarks as far as the density dependence is concerned. 
We use the factorization hypothesis for the four-quark condensate in this section. 
Another possible density dependence of the four-quark condensate and its effect to the result will be discussed 
in section \ref{Sec: discussion}. 
It is important to notice that the $\langle \overline{q} q \rangle$ term does not appear in 
the correlation function due to the structure of the interpolating field $J_{\Lambda_c}$. 
The di-quark structure in $J_{\Lambda_c}$ can be described as 
$\left (q^{Ta} C \gamma _5 q^{b} \right) = \left (- q^{Ta}_{L} C \gamma _5 q^{b}_{L} + q^{Ta}_{R} C \gamma _5 q^{b}_{R} \right)$
by using the left and the right handed spinors, and thus the quark condensate $\langle \overline{q} q \rangle$ where a 
left-handed spinor is paired to a right-handed one does not have contributions in the chiral limit. 
Thus $\Lambda_c$ feels the in-medium effects mainly through the four-quark condensate and $\langle q^{\dagger }q\rangle_m$. 

\subsection{Total energy}
\begin{figure*}[!tbp]
\begin{center}
  \includegraphics[width=12.cm]{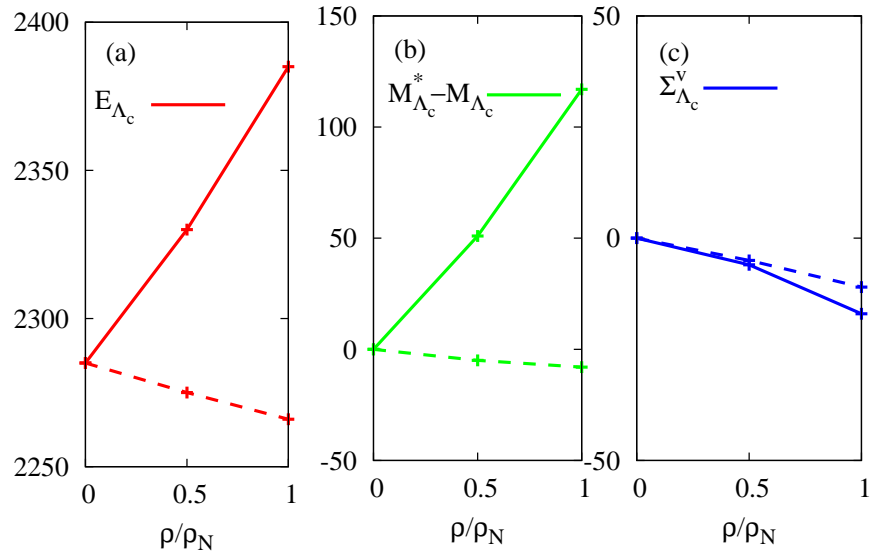}
 \end{center}
 \caption{The density dependences of (a) the energy $E_{\Lambda_c}$ [MeV], (b) the effective mass $M_{\Lambda_c}^{*}$ [MeV] and (c) the vector self-energy $\Sigma_{\Lambda_c}^{v}$[MeV]. 
The solid lines show the results with the density dependence of the four-quark condensate according to the factorization hypothesis while 
 dashed lines is the case where its density dependence are estimated from the perturbative chiral quark model.
}
 \label{fig: Charmed result}
\end{figure*}

We investigate the hadron properties by minimizing $\chi ^2$. 
In the case of the positive parity state, it is found that 
various combinations of the parameters lead the value of $\chi ^2$ to vicinity of 1. 
The trivial solutions are contained in these combinations.
The values that are close to $E_{\Lambda_c}=m_c$ and $q_{th}=m_{c}$ give small values of $\chi ^2 -1 $ because, except for the 
terms of the dimension 8 condensate and the part of the gluon condensate, 
$G_{mOPE} (\tau)$ can be expressed by $\delta (q_0 - m_c)$ and $\theta (q_0 - q_{th})$. 
These values do not correspond to physical solutions and must be 
discarded. 
In order to avoid such solutions, we impose further conditions. 
The interpolating operator $J_{\Lambda_c}$ for $\Lambda_c$ couples to 
the continuum states starting from around the value of the lowest threshold, $\Sigma_c \pi$, i.e., 2600 MeV. 
Therefore, we exclude the $q_{th}$ parameter region, $q_{th} < 2600 \ \mathrm{MeV}$. 
The $q_{th}$ dependence of the result in vacuum are summarized in Table\,\ref{tab: resultp in vacuum}. 
We find that the uncertainty of the calculated ground state mass is about 100 MeV and 
the choice of $q_{th} = 2.715 \ \mathrm{GeV}$ can reproduce the experimental value. 
\begin{table}[!t]
\begin{center}
\begin{tabular}{|c|c|c|c|c|c|c|c|}
\hline 
$q_{th}$ [GeV]               &2.6 &2.7        &2.715 &2.8    &2.9    &3.0     & 3.1 \\ \hline
$E_{\Lambda_c} $ [GeV] &2.218&2.275   &2.285 &2.320 &2.370  &2.397  & 2.419 \\ \hline
$|\lambda|^2$ [$10^{-4}\mathrm{GeV}^6$]              &1.20&1.55      &1.62  &1.97   &2.40   &2.77   & 3.08 \\ \hline
$\chi ^2$                     &1.002&1.007  &1.009 &1.020  &1.046 &1.078  & 1.116 \\ \hline
\end{tabular}
\caption{The $q_{th}$ dependence of the result.
}
\label{tab: resultp in vacuum}
\end{center}
\end{table}
Therefore, we will fix the value of $q_{th}$ as 2.715 GeV and 
apply the analyses to the $\Lambda_c$ baryon in nuclear matter. 
The density dependence of the energy $E_{\Lambda_c}$ is shown in Fig.\,\ref{fig: Charmed result}.  
We find that the energy of $\Lambda_c$ in nuclear matter, namely the peak position in the spectral function, increases as the density increases. 
At the normal nuclear matter density, its value is 2.385 GeV. 
This result indicates that $\Lambda_c$ feels a net repulsive potential in nuclear matter. 
The behavior is different from the nucleon and $\Lambda$, whose 
total energy gradually decreases and causes the formation of bound states in nuclei. 
As for the negative parity state, the analysis does not work well 
due to the small coupling between the present interpolating field and $\Lambda_c^{-}$ state. 

\subsection{Effective mass and vector self-energy}
The analyses in the previous subsection show that the energy of $\Lambda_c$ in nuclear matter increases. 
In relativistic phenomenological models, the total energy can be decomposed into the 
effective mass $m^{*}_{B}$ and the vector self-energy $\Sigma_{B}^{v}$ as we have introduced in Eq.\,(\ref{eq: eenrgy}). 
These quantities have been investigated from the previous QCD sum rule analyses for the 
nucleon \cite{Drukarev:1988kd,Drukarev:1991fs,Drukarev:2004zg,Drukarev:2009ac,Drukarev:2010xv,Drukarev:2012av,Cohen:1991js,Furnstahl:1992pi,Jin:1992id,Jin:1993up,Jin:1993fr} 
and $\Lambda$ \cite{Jin:1993fr,Cohen:1994wm,Kryshen:2011ng,Azizi:2015ica,Jeong:2016qlk}. 
The studies show that significant cancellations of the modifications of  $m^{*}_{B}$ and $\Sigma_{B}^{v}$ for both the nucleon 
and $\Lambda$ cases occurs. 
The modifications of $m^{*}_{B}$ and $\Sigma_{B}^{v}$ can 
be connected with the coupling strength of the baryon $B$ to the scalar and the vector mean fields, respectively. 
In a naive estimation where exchanged light mesons only couples to the light quarks in the baryon, 
one can derive a simple relation between the coupling strengths of the nucleon and that of $\Lambda$. 
If this approximation is valid, the effective masses and the vector self-energies satisfy the following relation: 
$M^{*}_{\Lambda} - M_{\Lambda} \approx \frac{2}{3} \left ( M^{*}_{N} - M_{N} \right ) $ and
$\Sigma_{\Lambda}^{v} \approx \frac{2}{3}\Sigma_{N}^{v}$. 
However, the previous QCD sum rule analyses show a large violation of the above relation \cite{Jin:1993fr,Cohen:1994wm,Kryshen:2011ng,Azizi:2015ica,Jeong:2016qlk}. 
We will investigate the effective mass and the vector self-energy of $\Lambda_c$ and compare 
the results with those of the nucleon and $\Lambda$.  

As we see from Eq.\,(\ref{eq: pnspf}), we can study the energy $E_{\Lambda_c^{\pm}}$ in the parity projected QCD sum rule. 
Here we parametrize the phenomenological side of each $G_{miOPE} (\tau) \ (i=1, 2, 3)$ and investigate the effective mass and the vector self-energy. 
The specific forms of phenomenological sides are as follows:  
\begin{equation}
\begin{split}
G_{m1\mathrm{SPF}} (\tau ) =& \int_{0}^{\infty} dq_{0} \frac{1}{\sqrt{4\pi \tau}} \exp \left ( -\frac{(q_0^2 - m_c^2)^2}{4 \tau} \right )  \\
                          & \times \Bigg (|\lambda ^+|^2 \frac{M_{\Lambda_c}}{2\sqrt{M^{*2}_{\Lambda_c}}} \delta (q_0 - E_{\Lambda_c} ) \\
                          & \hspace{1cm}  + \frac{1}{\pi} \mathrm{Im} \left [\Pi_{m1\mathrm{OPE}} (q_0) \right ] \theta (q_0 -q_{th}) \Bigg ) \\
G_{m2\mathrm{SPF}} (\tau ) &= \int_{0}^{\infty} dq_0 \frac{1}{\sqrt{4\pi \tau}} \exp \left ( -\frac{(q_0^2 - m_c^2)^2}{4 \tau} \right )  \\
                           & \times \Bigg ( |\lambda ^+|^2 \frac{M^*_{\Lambda_c}}{2\sqrt{M^{*2}_{\Lambda_c}}} \delta (q_0 - E_{\Lambda_c} ) \\
                           & \hspace{1cm} + \frac{1}{\pi} \mathrm{Im} \left [\Pi_{m2\mathrm{OPE}} (q_0) \right ] \theta (q_0 -q_{th})  \Bigg )\\
G_{m3\mathrm{SPF}} (\tau ) &= \int_{0}^{\infty} dq_0 \frac{1}{\sqrt{4\pi \tau}} \exp \left ( -\frac{(q_0^2 - m_c^2)^2}{4 \tau} \right ) \\
                           & \times \Bigg ( |\lambda ^+|^2 \frac{-\Sigma ^{v}_{\Lambda_c}}{2\sqrt{M^{*2}_{\Lambda_c}}} \delta (q_0 - E_{\Lambda_c} ) \\
                           & \hspace{1cm} + \frac{1}{\pi} \mathrm{Im} \left [\Pi_{m3\mathrm{OPE}} (q_0) \right ] \theta (q_0 -q_{th}) \Bigg ), 
\end{split}
\end{equation}
where the structure of the lowest lying pole comes from the the in-medium propagators of $\Lambda_c^{\pm}$ of Eq.\,(\ref{eq: in-medium propagator}). 
We do not consider the contribution of negative parity state $\Lambda_c^{-}$ in the above equation because 
its contribution are quite small. 
The validity of such treatment in vacuum are guaranteed by the fact that $G_{m\mathrm{OPE}}^{-} (\tau)$ is much 
smaller than the $G_{m\mathrm{OPE}}^{+} (\tau)$. 
Its validity in nuclear matter can be also investigated by comparing 
$G_{m2\mathrm{SPF}}(\tau )$ with $G_{m2\mathrm{OPE}} (\tau)$. 
$G_{m2\mathrm{SPF}}(\tau )$ is characterized by $q_{th}$, $|\lambda^{+}|^2$ and $E_{\Lambda_c}$ whose values
have been already obtained from the analyses of the parity projected QCD sum rules.   
We confirm that such treatment is valid up to the normal nuclear matter density. 
By fitting the $G_{m1\mathrm{SPF}} (\tau ) $, $G_{m3\mathrm{SPF}} (\tau )$ and corresponding functions in the OPE side, we investigate 
the effective mass and the vector self-energy. 
These density dependences are shown in Fig.\,\ref{fig: Charmed result}. 
From this figure, we find that the effective mass increases while the vector self-energy decreases as the density increases. 
At the normal nuclear matter density, the values of the effective mass and the vector self-energy 
is 2.402 GeV and -0.017 GeV, respectively. 
Comparing them with the previous $\Lambda_c$ QCD sum rule analysis with the same interpolating operator \cite{Wang:2011hta}, we agree in the increase of the effective mass 
but the sign of the vector self-energy is different. 
The discrepancy may come from the first order $\alpha_s$ corrections 
and the dimension 8 condensate. 
The small contribution of the dimension 8 condensate implies the good convergence of the OPE 
and thus we can use $G_{mOPE}(\tau)$ in lower energy region which contains much information about the $\Lambda_c$ ground state. 
In another preceding  study where a different interpolating operator is used \cite{Azizi:2016dmr}, 
both the signs of the modifications of the effective mass and the vector self-energies are opposite to our results and 
the magnitude of their values are much larger than ours. 

Comparing our results with the previous QCD sum rule of the nucleon\cite{Jin:1993fr} and $\Lambda$ \cite{Jeong:2016qlk}, we find that 
the shift of the effective mass is about $M^{*}_{\Lambda_c} - M_{\Lambda_c} \approx -(M^{*}_{\Lambda} - M_{\Lambda}) \approx -\frac{1}{3}(M^{*}_{N} - M_{N})$ 
and the vector self-energy is about $\Sigma_{\Lambda_c}^{v} \approx -\frac{1}{3} \Sigma_{\Lambda}^{v} \approx -\frac{1}{12} \Sigma_{N}^{v}$.
Their values of $\Lambda_c$ are quite different. 
The smallness of the vector self-energy may be understood from the contribution of the $\langle q^{\dagger} q \rangle$ term in $G_{m3\mathrm{OPE}} (\tau)$. 
In the case of the nucleon and $\Lambda$ analyses \cite{Cohen:1991js,Furnstahl:1992pi,Jin:1992id,Jin:1993up,Jin:1993fr}, 
this contribution plays a dominant role to determine the vector self-energy. 
The $\langle q^{\dagger} q \rangle$ contribution gives a repulsive vector self-energy of $\Lambda_c$ 
as well as of the nucleon and $\Lambda$. 
However, the contribution in $\Lambda_c$ is small 
because the Wilson coefficient of this term is proportional to $\frac{(q_0^2 - m_c^2)^4}{(q_0^2)^3}$ and its value 
becomes small as $q_0$ goes close to $m_c$. 
Therefore, when we investigate $\Lambda_c$ whose energy is close to charm quark mass, 
the contribution of $\langle q^{\dagger} q \rangle$ to the vector self-energy is suppressed. 
This suppression implies that the coupling strength of the vector field is affected by not only the light quarks 
but also the heavy quark because the Wilson coefficient contains both the 
light and the heavy quark propagators. 
 
\section{Discussion \label{Sec: discussion}}
We have investigated the density dependence of the energy, the effective mass and the vector self-energy 
of $\Lambda_c$. The results are different from our naive expectation and the qualitative behaviors 
of the nucleon and $\Lambda$. 
Here we discuss the density dependence of the four-quark condensate and 
its effects to the results of the analyses. 
As we have shown in Fig.\,\ref{fig: OPE density charm}, the four-quark condensate is dominant and thus 
will strongly affect the density dependence of $G_{mOPE}(\tau)$. 
The expectation value of the four quark condensate has large uncertainty and the deviation from factorization hypothesis
has been pointed out in previous studies \cite{Narison:1995jr,Bertlmann:1987ty,Dominguez:1987nw,Gimenez:1990vg,Kwon:2008vq,
Gimenez:1990vg,Gubler:2015yna,Chung:1984gr,Jin:1993fr,Celenza:1994ri,Thomas:2007gx,Drukarev:2012av}. 
Here, we use an effective model to estimate the density dependence of the four quark condensate and then 
reinvestigate the properties of $\Lambda_c$. 
To discuss the validity of the estimation of the four quark condensate, we apply the same QCD sum rule analyses to $\Lambda$ hyperon (
with appropriate strange quark condensates at finite density, see section IV B) 
and compare the results with the energy shift of $\Lambda$ in nuclear matter. We also investigate the in-medium modification of $\Lambda_b$. 

\subsection{Density dependence of four-quark condensate}
In this subsection, we discuss the density dependence of the four-quark condensate and its effect to 
the in-medium modifications of  $\Lambda_c$. 
We specifically consider two types of density dependences. 
The first one is based on the factorization hypothesis and the second is estimated from the perturbative chiral quark model (PCQM) \cite{Drukarev:2003xd,Thomas:2007gx}. 

The in-medium condensates are usually evaluated in the linear density approximation  
and can be described as $\langle \mathcal{O} \rangle _{m} \cong  \langle \mathcal{O} \rangle _{0} + \rho \langle \mathcal{O} \rangle _{N}$.   
It is important to obtain precise values of $\langle \mathcal{O} \rangle _{N}$, the values of the condensates in the nucleon. 
In some cases, they can be experimentally subtracted. For instance, expectation values of some twist-4 four-quark condensates in the nucleon can be 
estimated from deep inelastic scattering data \cite{Choi:1993cu,Jeong:2012pa,Jeong:2016qlk}. 
However, general four-quark condensates can not be determined. 
A commonly used technique is the factorization hypothesis, where a four-quark condensate is given by the product of two quark condensates as 
\begin{equation}
\begin{split}
\langle u^{a}_{\alpha} \overline{u}^{b}_{\beta} d^{c}_{\gamma} \overline{d}^{d}_{\delta} \rangle &\rightarrow 
\langle u^{a}_{\alpha} \overline{u}^{b}_{\beta} \rangle \langle d^{c}_{\gamma} \overline{d}^{d}_{\delta} \rangle, 
\label{eq: factorization}
\end{split}
\end{equation}
where $a$, $b$, $c$ and $d$ are color indices and $\alpha, \beta, \gamma$ and $\delta$ are spinor indices.  
The factorization can only be justified in the large $N_c$ limit \cite{Shifman:1978bx,Shifman:1978by} and 
the validity in $N_c=3$ is not so clear.  
In fact, there are some studies which claim significant violation of the 
factorization in the QCD sum rule analyses of $\rho$ meson \cite{Narison:1995jr,Bertlmann:1987ty,Dominguez:1987nw,Gimenez:1990vg,Kwon:2008vq}, 
$\phi$ meson\cite{Gimenez:1990vg,Gubler:2015yna} and the nucleon \cite{Chung:1984gr,Jin:1993fr,Celenza:1994ri,Thomas:2007gx,Drukarev:2012av}. 
Deviation from the factorization in the $\Lambda_c$ channel and its affect to the results of the analyses have not been studied. 

We here consider a model dependent but more sophisticated approach based on PCQM. 
In general, the four-quark condensates which appear in the OPE calculations have different density dependences. 
The four-quark condensates can be expressed as linear combinations of the independent four-quark condensates 
and their density dependence are different from each other.  
In Refs.\,\cite{Drukarev:2003xd,Thomas:2007gx}, the nucleon expectation values of the various independent four-quark condensates 
are estimated by PCQM. 
In the case of the nucleon channel, the in-medium four-quark condensates which are estimated by PCQM 
can quantitatively reproduce the known properties of the nucleon in nuclear matter \cite{Thomas:2007gx}. 
We estimate the density dependence of the four-quark condensate 
from these results and reinvestigate the in-medium modification of $\Lambda_c$. 

In the case of the interpolating operator $J_{\Lambda_c}$, the structure of the four-quark condensate can be described as 
\begin{equation}
\begin{split}
& \big \langle \left ( \epsilon ^{abc} u ^{a} C\gamma_5  d^{b} \right ) \cdot
\left ( \epsilon^{efc} \overline{d} ^{f} \gamma_5 C \overline{u}^{e} \right )  \big \rangle_m \\
&= -\frac{1}{4}  \Bigg[ \langle \overline{d} ^{f} d^{b} \overline{u} ^{e} u^{a} \rangle_{m} + \langle \overline{d} ^{f} \gamma _{5} d^{b} \overline{u} ^{e} \gamma _{5} u^{a} \rangle_{m} \\
&\hspace{1.5cm} -\frac{1}{2} \langle \overline{d} ^{f} \sigma _{\mu \nu} d^{b} \overline{u} ^{e} \sigma ^{\mu \nu} u^{a} \rangle_{m}
                        +  \langle \overline{d} ^{f} \gamma _{\mu} d^{b} \overline{u} ^{e} \gamma ^{\mu} u^{a} \rangle_{m} \\
&\hspace{1.5cm} + \langle \overline{d} ^{f} \gamma_{5} \gamma _{\mu} d^{b} \overline{u} ^{e} \gamma_5 \gamma ^{\mu} u^{a} \rangle_{m}   \Bigg ] \epsilon ^{abc} \epsilon ^{efc}. 
\label{eq: four quark}
\end{split}
\end{equation}
Here, we have decomposed the four-quark condensate into independent four-quark condensates 
which have the different Lorentz structures. 
We find that the twist-0 four-quark condensates only appear in this sum rule. 
In the case of the factorization hypothesis, the density dependence of the 
four-quark condensate is expressed as 
\begin{equation}
\begin{split}
&\big \langle \left ( \epsilon ^{abc} u ^{a} C\gamma_5  d^{b} \right ) \cdot
\left ( \epsilon^{efc} \overline{d}^{f} \gamma_5 C \overline{u}^{e} \right ) \big \rangle _{m} \\
&= -\frac{1}{6} \left ( \langle \overline{d}d \rangle _{m}  \langle \overline{u}u \rangle _{m} + \langle d^{\dagger} d \rangle _{m}  \langle u^{\dagger} u \rangle _{m} \right ) \\
&= -\frac{1}{6} \left ( \langle \overline{q}q \rangle _{m}  ^2 + \langle q^{\dagger} q \rangle _{m}  ^2 \right ) \\
&= -\frac{1}{6} \left (\langle \overline{q}q \rangle _{0}  ^2 + \rho \frac{\sigma_N}{m_q} \langle \overline{q}q \rangle _{0} + (\frac{\sigma_N^2}{4m_q^2}+\frac{9}{4}) \rho^2 \right ), 
\label{eq: 4quark fcatorization}
\end{split}
\end{equation}
where the isospin symmetry is used. 
This density dependence is used in the previous sections. 
On the other hand, from the results of the PCQM calculations \cite{Drukarev:2003xd,Thomas:2007gx}, 
the density dependence of whole four-quark condensate can be written as 
\begin{equation}
\begin{split}
&\big \langle \left ( \epsilon ^{abc} u ^{a} C\gamma_5 d ^{b} \right ) \cdot
\left ( \epsilon^{efg} \overline{d} ^{f} \gamma_5 C \overline{u}^{e} \right ) \big \rangle _{m} \\
&= -\frac{1}{6} \langle \overline{q}q \rangle _{0}  ^2  -\frac{1}{4}  0.935 \langle \overline{q} q \rangle _{0}  \ \rho 
+ \mathcal{O} (\rho^2). 
\label{eq: 4quark PCQM}
\end{split}
\end{equation}
The coefficient of $\langle \overline{q}q \rangle _{0}  ^2$ is determined by the factorization hypothesis. 
For simplicity, we call the density dependences of Eq.\,(\ref{eq: 4quark fcatorization}) and Eq.\,(\ref{eq: 4quark PCQM}) 
as F-type and QM-type, respectively. 
Comparing Eq.\,(\ref{eq: 4quark fcatorization}) with Eq.\,(\ref{eq: 4quark PCQM}), 
we find that the QM-type in-medium modification is much weaker than that that of the F-type. 
Using this density dependence, we recalculate the in-medium modifications of $\Lambda_c$. 
The results are shown as dashed lines in Fig.\,\ref{fig: Charmed result}. 
At the normal nuclear matter density, the values of the energy, the effective mass and the vector self-energy 
are 2.266 GeV, 2.277 GeV and -0.011 GeV, respectively. 
The total energy decreases about 20 MeV, which implies that $\Lambda_c$ feels a net attractive potential and 
forms bound states in nuclear matter.
The results indicate that the density dependence of the four-quark condensate
strongly affects the in-medium modifications of $\Lambda_c$. 
Therefore, in turn, $\Lambda_c$ is useful as a probe of the density dependence of the four-quark condensate. 

Finally, we comment on the relation between the partial restoration of chiral symmetry and the 
four-quark condensate of Eq.\,(\ref{eq: four quark}). 
The chiral condensate is usually considered as an order parameter of the chiral transition, 
but the role of four-quark condensates in the spontaneous breaking of the chiral symmetry is still an open issue. 
The effects from four-quark condensates to the phase transition are discussed in Refs.\,\cite{Heinz:2008cv,Gallas:2011qp,Mukherjee:2013jca,Pisarski:2016ukx}. 
However, the four-quark condensate of Eq.\,(\ref{eq: four quark}) is singlet under the chiral $SU(2)\times SU(2)$ transformation and thus its in-medium modification 
is not directly related to the partial restoration of chiral symmetry. 
The knowledge of this density dependence may be useful when we investigate other hadrons in nuclear matter. 
Some kinds of four-quark condensates which appear in OPE may contain the four quark condensate of Eq.\,(\ref{eq: four quark}). 

\subsection{Analyses of $\Lambda$ and $\Lambda_b$}
We carry out the analyses of $\Lambda$ to discuss the validity of the estimation of the in-medium modification of the four quark condensate 
in this subsection. 
The OPE of $\Lambda$ contains some 
extra terms in which the strange quark condensate is treated differently from the charm quark. 
The Wilson coefficients of the gluon condensate and strange quark condensate terms 
are different from the case of $\Lambda_c$. 
Therefore, we refer Ref.\,\cite{Jeong:2016qlk} and 
construct the parity projected Gaussian sum rule of $\Lambda$. 
We also investigate the properties of $\Lambda_b$ in nuclear matter. 
The OPE representation of the $\Lambda_b$ correlation function is same as that of $\Lambda_c$ 
except for the value of the quark mass $m_Q$. 
The values of the new parameters are as follows: 
$m_b=4.78\pm0.06$ GeV \cite{Agashe:2014kda}, $m_s=130\pm8$ MeV \cite{Agashe:2014kda}, 
$\langle \overline{s}s\rangle _{0}=0.8 \ \langle \overline{q}q\rangle _{0}$ \cite{Jeong:2016qlk} and 
$\langle \overline{s}s\rangle _{N} = 0.1 \ \langle \overline{q}q\rangle _{N}$ \cite{Jeong:2016qlk}. 
The analyzed parameter region for $\Lambda$ and $\Lambda_b$ are $0.6<\tau[\mathrm{GeV}^4]<1.8$ and 
$9.5<\tau[\mathrm{GeV}^4]<23$, respectively, which are determined by the same criterion as the $\Lambda_c$ analyses.  
\begin{figure}[!tbp]
\begin{center}
  \includegraphics[width=9.cm]{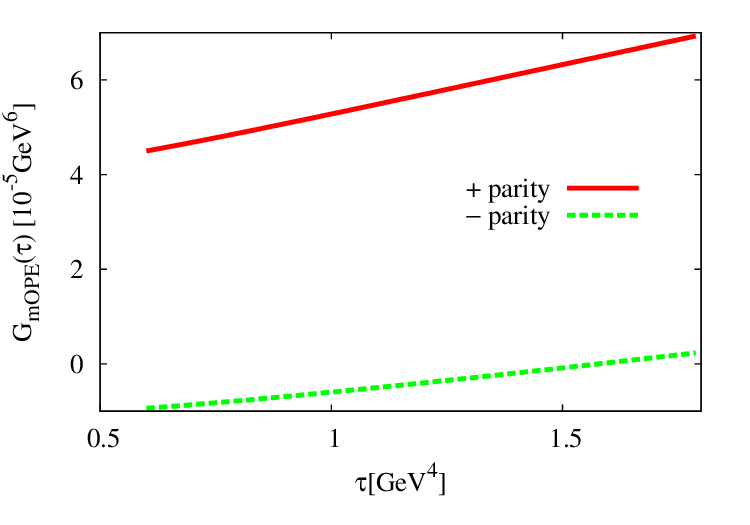}
 \end{center}
 \caption{The positive and the negative parity OPE $G^{\pm}_{m\mathrm{OPE}} (\tau)$ of $\Lambda$ in vacuum. 
}
 \label{fig:OPE p and n strange}
\end{figure}

\begin{figure}[!tbp]
\begin{center}
  \includegraphics[width=9.cm]{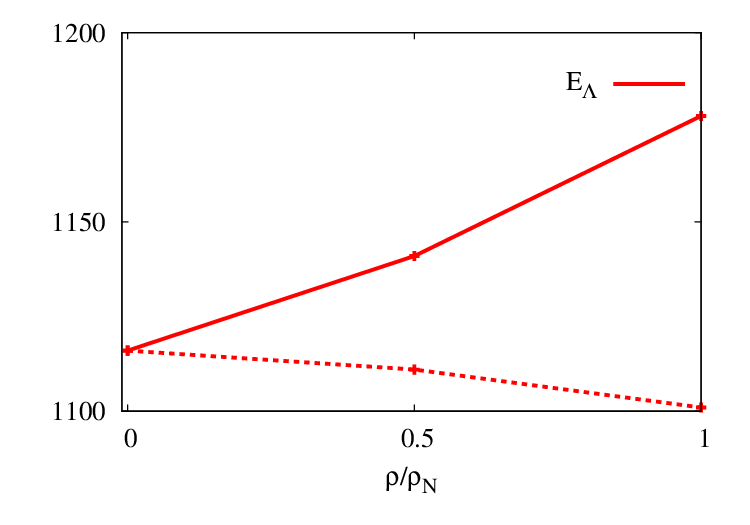}
 \end{center}
 \caption{The density dependences of the energy $E_{\Lambda}$ [MeV]. 
The solid line shows the result with the density dependence of the four-quark condensate according to the factorization hypothesis while the 
 dashed line is the case where its density dependence is estimated from the perturbative chiral quark model.
}
 \label{fig: hyperon result}
\end{figure}

\begin{figure*}[!tbp]
\begin{center}
  \includegraphics[width=12.cm]{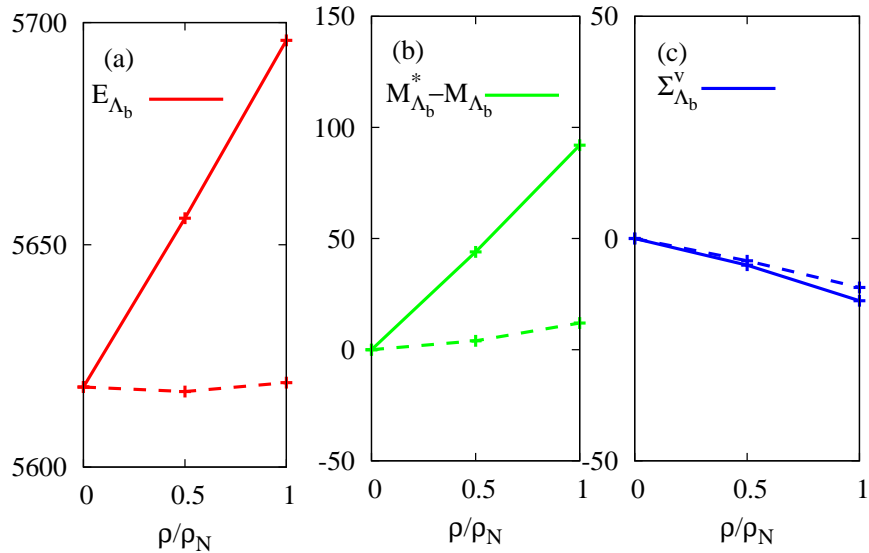}
 \end{center}
 \caption{The density dependences of (a) the energy $E_{\Lambda_b}$ [MeV], (b) the effective mass $M_{\Lambda_b}^{*}$ [MeV] and (c) the vector self-energy $\Sigma_{\Lambda_b}^{v}$[MeV]. 
The solid lines show the results with the density dependence of the four-quark condensate according to the factorization hypothesis while 
 dashed lines is the case where its density dependence are estimated from the chiral quark model.
}
 \label{fig: Bottomed result}
\end{figure*}

The behavior of  the OPE for $\Lambda$ is shown in Fig.\,\ref{fig:OPE p and n strange}. 
We find that the interpolating operator couples to both the positive and negative parity states, in contrast to the case of $\Lambda_c$. 
The result indicates that we should take into account effects of the negative parity state 
when investigating the effective mass and the vector self-energy of the positive parity state.  
Therefore, we leave the individual quantities for a future work and investigate the density dependence only of the energy of the positive parity state in this study. 
The qualitative behavior of $G_{mOPE} (\tau)$ of $\Lambda_b$ is same as that of $\Lambda_c$. 
The results of the analyses of $\Lambda$ and $\Lambda_b$ are shown in Figs. \ref{fig: hyperon result} and \ref{fig: Bottomed result}. 
The values of the threshold parameter is fixed to $q_{th}^{\Lambda} = 1.52 \ \mathrm{GeV}$ and $q_{th}^{\Lambda_b} = 6.03 \ \mathrm{GeV}$, respectively. 
They are taken so as to reproduce the experimental mass in vacuum. 

Fig.\,\ref{fig: hyperon result} shows 
that energy of $\Lambda$ increases in the case of the F-type in-medium four-quark condensate 
while the energy slightly decreases when we use the QM-type in-medium four-quark condensate. 
The energy shift of $\Lambda$ in nuclear matter has been extracted from the binding energies of hypernuclei (see Ref.\,\cite{Hashimoto:2006aw} for a review), 
which is consistent with our result of the QM-type density dependence. 
This consistency supports that the QM-type density dependence is more realistic than the F-type, 
namely the density dependence of the four-quark condensate of Eq.\,(\ref{eq: four quark}) is quite weak. 
The results support that $\Lambda_c$ in nuclear matter feels an weak attractive force. 
As for the $\Lambda_b$ analyses with the QM-type density dependence, 
the energy is almost independent of the density, which implies the difficulty of forming a bound state. 
Comparing the results of $\Lambda_b$ with those of $\Lambda_c$, we find that the values of 
the energy shift, the effective mass shift and the vector self-energy of  $\Lambda_b$ and those of $\Lambda_c$ are of the same scale. 
These behaviors come from the small in-medium modifications of 
the four-quark condensate and $\langle q^{\dagger}q \rangle$. 

\section{Summary and conclusion \label{Sec: Summary}}
We have studied the properties of $\Lambda_c$ in nuclear matter by using the QCD sum rule. 
To improve the $\Lambda_c$ QCD sum rule, 
we have taken into account the first order $\alpha_s$ correction 
and the higher order condensate terms in the OPE and then construct the 
parity projected QCD sum rule. 
The employed $\Lambda_c$ interpolating operator is the scalar di-quark-heavy quark type 
whose structure is same as the quark model picture of the $\Lambda_c$ ground state. 
In the OPE side, the four-quark condensate is dominant in vacuum, 
as $\langle \overline{q} q \rangle$ does not appear due to the structure of the interpolating operator. 
Therefore, $\Lambda_c$ feels the in-medium effects mainly through the four-quark condensate. 
We find that our interpolating operator strongly couples to the positive parity state while the coupling to the negative parity state is weak. 

From the analysis of $\Lambda_c$, the density dependences of 
the energy $E_{\Lambda_c}$, the effective mass $m_{\Lambda_c}^{*}$ and the vector self-energy $\Sigma_{\Lambda_c}^{v}$ 
are obtained. 
We have found that the results depend strongly on the density dependence of the four-quark condensate. 
We have considered two cases. 
The first one is the density dependence according to the factorization hypothesis (F-type) and the second one is estimated from 
perturbative chiral quark model (QM-type). 
The density dependence of the QM-type is much weaker than that of the F-type. 
The result with the F-type in-medium four-quark condensate 
shows that the energy and the effective mass increase while the vector self energy decreases in nuclear matter. 
On the other hand, the result using the QM-type is that the energy, the effective mass and the vector self-energy decrease in nuclear matter, 
which indicates the $\Lambda_c$ bound states in nuclear matter. 
The obtained binding energy is about 20 MeV at the normal nuclear matter density. 
The sensitivity to the in-medium modification 
of four-quark condensate implies that $\Lambda_c$ is useful as a probe of the density dependence of four-quark condensate. 
Comparing the values of $m_{\Lambda_c}^{*}$ and $\Sigma_{\Lambda_c}^{v}$ with those of $\Lambda$ in the previous QCD sum rule analyses, 
we find that their values are considerably different from each other. 
This implies the large violation of the naive expectation where exchanged light mesons only coupled to light quarks in the baryon. 
From the viewpoint of the OPE expression, the discrepancy can be understood as follows. 
The smallness of the vector self-energy come from the $\langle q^{\dagger} q \rangle$ term 
whose contribution plays a dominant role to determine the vector self-energy in the case of the nucleon and $\Lambda$.   
The Wilson coefficient of this term becomes small as $q_0$ goes close to $m_Q$. 
Due to this property, when we investigate $\Lambda_Q$ whose energy is close to the heavy quark mass, 
the contribution of $\langle q^{\dagger} q \rangle$ to the vector self-energy is suppressed. 
As a result, the value of $\Sigma_{\Lambda_Q}^{v}$ becomes small. 
For the effective mass, its value is also small as there is no contribution from the quark condensate $\langle \overline{q}q \rangle$. 

We have applied the parity projected QCD sum rule to the analyses of $\Lambda$ and $\Lambda_b$ in nuclear matter. 
It is found that the in-medium modification of the energy of $\Lambda$ depends on the density dependence of the four-quark condensates and 
the result with the QM-type density dependence is qualitatively consistent with the experimental results. 
This consistency supports that 
the density dependence  of the four-quark condensate is quite weak. 
Therefore, we conclude that $\Lambda_c$ in nuclear matter feels weak attraction. 

The results of the $\Lambda_b$ analyses using the QM-type density dependence show that the energy is almost density independent, 
which implies the difficulty of forming bound states. 
Comparing the results of $\Lambda_b$ with those of $\Lambda_c$, we find that the values of 
the energy, the effective mass and the vector self-energy of  $\Lambda_c$ and those of $\Lambda_b$ are of the same scale. 
These behaviors come from the small in-medium modification of 
the four-quark condensate and $\langle q^{\dagger}q \rangle$. 

\begin{acknowledgments}
The authors gratefully thank Kei Suzuki and Shigehiro Yasui for fruitful discussions. 
K.J.A. was supported by Grant-in-Aid for JSPS
Fellows from Japan Society for the Promotion of Science
(JSPS) (No.15J11897). 
This work is supported by Grants-in-Aid for Scientific Research from JSPS [Grant No. JP25247036(A)]. 
\end{acknowledgments}

\end{document}